\documentstyle[12pt,a4]{article}

\begin{document}

\begin{titlepage}
\begin{flushright}
{\large \bf UCL-IPT-97-07}
\end{flushright}
\vskip 2cm
\begin{center}

{\Large \bf
A Dynamical Scheme for a Large $CP$-Violating Phase}
\vskip 1cm

{\large D. Del\'epine,  J.-M. G\'erard,  R. Gonz\'alez Felipe and  J. Weyers}\\
\vskip 5 mm
{\em Institut de Physique Th\'eorique, Universit\'e catholique de
Louvain,}\\

{\em  B-1348 Louvain-la-Neuve, Belgium}

\end{center}

\vskip 2cm

\begin{abstract}
A dynamical scheme where the third generation of quarks plays a distinctive
 role is implemented. New interactions with a $\theta$ term induce the 
breaking of the electroweak symmetry and the top-bottom mass splitting. 
A large $CP$-violating phase naturally follows from the latter.

\end{abstract}

\end{titlepage}

\medskip

\section{Introduction}

\medskip

Among the outstanding problems in particle physics are the mechanism for the
electroweak symmetry breaking and the closely related question of the origin
of the quark masses. The fact that the top quark is very heavy compared to
the other quarks, $m_{t}=180\pm 12$ GeV \cite{PDG}, suggests that the third
generation may be playing a special role in the dynamics at the electroweak
scale. In particular, new strong interactions might exist at this scale
which not only distinguish the third generation from the others but also
intimately participate in the breaking of the electroweak symmetry. To
implement this scenario one may consider effective four-fermion
interactions \cite{NJL}. They can lead to the formation of quark-antiquark
bound states which in turn trigger dynamically the breaking of the
electroweak symmetry \cite{miransky,bardeen}.

More than thirty years after its discovery in the $K^{0}-\bar{K}^{0}$
system, it is fair to say that the mechanism for $CP$ violation is not yet
understood. But there is little doubt that $CP$ violation is deeply rooted
in the peculiar mass spectrum of the quarks. Nowadays, the familiar
puzzles related to an unexpectedly heavy top quark and to small mixing
angles in the Cabibbo-Kobayashi-Maskawa (CKM) matrix \cite{kobayashi} have
been turned around into questions concerning tiny quark masses and a large 
$CP$-violating phase. The main purpose of this paper is to suggest a
dynamical framework where these ``new'' points of view are connected.

Two approaches are usually considered when introducing $CP$ violation into 
gauge models: $CP$-violating phases can be explicit in the
parameters of the theory or alternatively, they can be spontaneously
generated through complex vacuum expectation values (VEVs) of the scalar
Higgs fields. In the standard model with complex parameters in the
Lagrangian, only two phases remain physical after field redefinitions,
namely, the QCD-induced phase $\theta _{QCD}$ and the phase $\delta $ in the
CKM matrix. The QCD phase will not concern us here \cite{peccei}.

In this paper, our starting assumption will be that the third generation of
quarks does indeed experience new forces (symmetric in $t$ and $b$) and that
these new forces also generate a $\theta $ term. We then propose a model
where this $\theta $ term breaks the symmetry between $t$ and $b$ and
induces naturally a large $\delta $ phase due to the smallness of the $%
m_{b}/m_{t}$ mass ratio.

\section{The model}

We consider a standard model Higgs sector in combination with an effective
new strong interaction acting on the third generation of quarks and
characterized by a $\theta$ term \cite{buchalla}. We require this new
strong interaction to conserve the isospin symmetry between $t$ and $b$
quarks.

Since the electroweak symmetry breaking will eventually be induced by 
radiative corrections \cite{fatelo} due to top-quark (and possibly, 
bottom-quark) loops, we may as well neglect the quartic self-interactions
 of the Higgs field. In this case, the relevant classical Lagrangian for
 the fundamental scalar field $H$ is given by

\begin{equation}
L_{H}=D_{\mu }H^{\dagger }D^{\mu }H-m_{H}^{2}H^{\dagger }H+\left( h_{t}\bar{%
\psi}_{L}t_{R}H+h_{b}\bar{\psi}_{L}b_{R}\tilde{H}\ +\mbox{ h.c.}\right) ,
\label{1}
\end{equation}
\smallskip where $H=\left( 
\begin{array}{l}
H^{0} \\ 
H^{-}
\end{array}
\right) $ , $\tilde{H}=\left( 
\begin{array}{l}
\;H^{+} \\ 
-H^{0^{*}}
\end{array}
\right) $ and $\psi _{L}=\left( 
\begin{array}{l}
t_{L} \\ 
b_{L}
\end{array}
\right) $; $h_{t}$ and $h_{b}$ are the Yukawa couplings and $D_{\mu }$ is
the usual covariant derivative of the standard model.

Next we assume that the interactions acting on the members of the third
generation of quarks are strong enough to form $q\bar{q}$ bound states at
the electroweak scale. We shall describe the latter in terms of two complex
doublet scalar fields

\begin{equation}
\Sigma _{t}=\left( 
\begin{array}{l}
\Sigma _{t}^{0} \\ 
\Sigma _{t}^{-}
\end{array}
\right) \sim t_{R}\bar{\psi}_{L}\;,\qquad \tilde{\Sigma}_{b}=\left( 
\begin{array}{l}
\;\Sigma _{b}^{+} \\ 
-\Sigma _{b}^{0^{*}}
\end{array}
\right) \sim b_{R}\bar{\psi}_{L}  \label{2}
\end{equation}
and the corresponding effective Lagrangian then reads:
\begin{equation}
L_{\Sigma }=D_{\mu }\Sigma _{t}^{\dagger }D^{\mu }\Sigma _{t}+D_{\mu }\Sigma
_{b}^{\dagger }D^{\mu }\Sigma _{b}-m^{2}(\Sigma _{t}^{\dagger }\Sigma
_{t}+\Sigma _{b}^{\dagger }\Sigma _{b})+g(\bar{\psi}_{L}t_{R}\Sigma _{t}+%
\bar{\psi}_{L}b_{R}\tilde{\Sigma}_{b}+\mbox{h.c.})\ .  \label{3}
\end{equation}

From Eqs.(\ref{1}) and (\ref{3}) it follows that the top and bottom
field-dependent masses are given at the tree level by the linear 
combinations

\begin{equation}
M_{t}=h_{t}H^{0}+g\Sigma _{t}^{0},\qquad M_{b}=h_{b}\tilde{H}^{0}+g\tilde{%
\Sigma}_{b}^{0},  \label{4}
\end{equation}
respectively (at this stage, mixing with the two light generations are of 
course neglected).

In the framework of this effective theory, effects of a (new) $\theta $ 
term can, in principle, be described via an arbitrary function of $\det U$, 
where

\begin{equation}
U\sim \left( 
\begin{array}{ll}
\bar{t}_{L}t_{R} & \bar{t}_{L}b_{R} \\ 
\bar{b}_{L}t_{R} & \bar{b}_{L}b_{R}
\end{array}
\right) ,  \label{5}
\end{equation}
or in terms of the composite fields defined in Eq.(\ref{2})

\begin{equation}
U=\left( 
\begin{array}{ll}
\Sigma _{t}^{0} & \;\Sigma _{b}^{-} \\ 
\Sigma _{t}^{+} & -\Sigma _{b}^{0^{*}}
\end{array}
\right) .  \label{6}
\end{equation}

\medskip In analogy with QCD $\cite{witten},$ we shall take the Lagrangian
form

\begin{equation}
L_{\theta }=-\frac{\alpha }{4}\left[ i\mbox{Tr }\left( \ln U-\ln U^{\dagger
}\right) +2\theta \right] ^{2},  \label{7}
\end{equation}
which typically arises as a leading term in a $1/N$ - expansion.

The total effective Lagrangian of our model is thus given by

\begin{equation}
L=L_{H}+L_{\Sigma }+L_{\theta }\;,  \label{8}
\end{equation}
with $L_{H},L_{\Sigma }$ and $L_{\theta }$ defined in Eqs.(\ref{1}),(\ref{3}%
) and (\ref{7}), respectively. We notice that if $h_{t}=h_{b}$ the total
Lagrangian (\ref{8}) conserves an ``isospin'' symmetry. As we shall see 
below, the $\theta $ angle will provide a dynamical origin for both $CP$ 
violation and isospin breaking \cite{vafa}, once the neutral components of 
the three doublets $H,$ $\Sigma _{b}$ and $\Sigma _{t}$ acquire nonzero VEVs.

\medskip

\section{Electroweak and isospin symmetry breakings}

\medskip Let us now discuss how the electroweak symmetry breaking and $CP$
violation arise in the present model. Without loss of generality, we take
the phase of the neutral Higgs field $H^{0}$ to be zero. This can always be
achieved by performing a suitable electroweak gauge transformation. We write
the VEVs of the neutral components of the fields in the form

\begin{equation}
\left\langle H^{0}\right\rangle =\frac{v}{\sqrt{2}}\;,\quad \left\langle
\Sigma _{t}^{0}\right\rangle =\frac{\sigma _{t}}{\sqrt{2}}e^{i\varphi
_{t}},\quad \left\langle \Sigma _{b}^{0}\right\rangle =\frac{\sigma _{b}}{%
\sqrt{2}}e^{i\varphi _{b}}.  \label{9}
\end{equation}

Including the radiative corrections (induced by top and bottom quark loops),
the effective potential in terms of these VEVs reads

\begin{equation}
V=m_{H}^{2}\frac{v^{2}}{2}+\frac{m^{2}}{2}(\sigma _{t}^{2}+\sigma
_{b}^{2})-\beta \left( \mu _{t}^{2}+\mu _{b}^{2}\right) +\lambda \left( \mu
_{t}^{4}+\mu _{b}^{4}\right) +\alpha \left( \theta -\varphi _{t}+\varphi
_{b}\right) ^{2}\;,  \label{10}
\end{equation}
\newline
\noindent where
\begin{eqnarray}
\mu _{t}^{2} &=&|\left\langle M_{t}\right\rangle |^{2}=\frac{1}{2}\left(
h_{t}^{2}v^{2}+g^{2}\sigma _{t}^{2}+2h_{t}vg\sigma _{t}\cos \varphi
_{t}\right) ,  \nonumber \\
\mu _{b}^{2} &=&|\left\langle M_{b}\right\rangle |^{2}=\frac{1}{2}\left(
h_{b}^{2}v^{2}+g^{2}\sigma _{b}^{2}+2h_{b}vg\sigma _{b}\cos \varphi
_{b}\right) ,  \label{11}
\end{eqnarray}
while $\beta $ and $\lambda $ are some effective quadratic and quartic 
couplings. In what follows we shall assume all couplings and parameters in 
the potential to be real and positive.

Note that the potential which leads to Eq.(\ref{10}) can be viewed
either as an effective renormalizable interaction or as an expansion up to
quartic terms in a cut-off theory.

The extrema conditions $\frac{\partial V}{\partial v}=\frac{\partial V}{%
\partial \sigma _{t}}=\frac{\partial V}{\partial \sigma _{b}}=\frac{\partial
V}{\partial \varphi _{t}}=\frac{\partial V}{\partial \varphi _{b}}=0$ imply
the following system of equations
\begin{eqnarray}
A_{H}v &=&gh_{t}I_{t}\sigma _{t}\cos \varphi _{t}+gh_{b}I_{b}\sigma _{b}\cos
\varphi _{b}\;,  \nonumber \\
A_{t}\sigma _{t} &=&gh_{t}I_{t}v\cos \varphi _{t}\ ,  \nonumber \\
A_{b}\sigma _{b} &=&gh_{b}I_{b}^{{}}v\cos \varphi _{b}\ ,  \nonumber \\
gh_{t}I_{t}v\sigma _{t}\sin \varphi _{t} &=&-gh_{b}I_{b}^{{}}v\sigma
_{b}\sin \varphi _{b}=2\alpha \left( \theta -\varphi _{t}+\varphi
_{b}\right) ,  \label{12}
\end{eqnarray}
\newline
where ($i=t,b$)
\begin{eqnarray}
A_{H} &=&m_{H}^{2}-h_{t}^{2}I_{t}-h_{b}^{2}I_{b}\;,  \nonumber \\
A_{i} &=&m^{2}-g^{2}I_{i}\;,  \nonumber \\
I_{i} &=&\beta -2\lambda \mu _{i}^{2}.  \label{13}
\end{eqnarray}
Notice the similarity of the last equation in Eq.(\ref{12}) with the one
appearing in QCD \cite{witten}. To Eqs.(\ref{11})-(\ref{13}) we should add
the normalization condition

\begin{equation}
v_{0}=\sqrt{v^{2}+\sigma _{t}^{2}+\sigma _{b}^{2}}\ =246\ {\rm GeV}\ ,
\label{14}
\end{equation}
coming from the $W$ boson mass.

We shall take the mass parameters $m_{H}$ and $m$ to be such that
\begin{equation}
m_{H}^{2}-(h_{t}^{2}+h_{b}^{2})\beta >0\;,\;m^{2}-g^{2}\beta >0.  \label{15}
\end{equation}
In this case $A_{H},$ $A_{t}$ and $A_{b}$ defined in (\ref{13}) are 
always positive.

From Eqs.(\ref{12})-(\ref{13}) follows the ``gap'' equation

\begin{equation}
\frac{A_{t}A_{H}}{h_{t}^{2}g^{2}I_{t}^{2}}=\cos ^{2}\varphi _{t}+\eta \cos
^{2}\varphi _{b}\equiv r\;,  \label{16}
\end{equation}
where
\begin{equation}
\eta =\frac{A_{t}h_{b}^{2}I_{b}^{2}}{A_{b}h_{t}^{2}I_{t}^{2}}\;.  \label{17}
\end{equation}
Moreover,

\begin{equation}
\sin ^{2}\varphi _{t}=\frac{\sigma _{b}^{2}(1-r+\eta )}{\eta \sigma
_{t}^{2}I_{t}^{2}/I_{b}^{2}+\sigma _{b}^{2}}\;,\;\sin ^{2}\varphi _{b}=\frac{%
\sigma _{t}^{2}(1-r+\eta )}{\eta \sigma _{t}^{2}+\sigma
_{b}^{2}I_{b}^{2}/I_{t}^{2}}\;  \label{18}
\end{equation}
and
\begin{equation}
\sin 2\varphi _{t}=-\eta \sin 2\varphi _{b}\;.  \label{19}
\end{equation}

If the parameter $\alpha $ is large ($\alpha \gg \beta
m_{t}^{2}$ with $m_{t}$ the physical mass of the top quark) the last
equation in (\ref{12}) implies the constraint

\begin{equation}
\theta \simeq \varphi _{t}-\varphi _{b}\;.  \label{20}
\end{equation}
Furthermore, if $\theta =0$, it is easily seen that $\varphi
_{t}=\varphi _{b}=0$ is the only solution of our equations and therefore $CP$
is conserved. Hence, $CP$ violation in the context of our model requires $%
\theta $ to be non zero.

Notice also that with $h_{b}=0$ and $h_{t}\neq 0$ the solution to Eqs.(\ref
{12}) is $\sigma _{b}=0$, $\theta =\varphi _{t}=0.$ In this case $m_{b}=0$,
the composite field $\Sigma _{b}$ is superfluous and we have a $CP$%
-conserving model with an elementary Higgs boson and a top-quark 
condensate \cite{delepine}.

We now proceed to solve Eqs.(\ref{16})-(\ref{19}) for the isopin
symmetric case $h_{t}=h_{b}\equiv h\neq 0.$

To illustrate a particularly simple analytical solution, let us
assume that $\beta \gg 2\lambda m_{t}^{2}$. Then (cf. Eq.(\ref{13})) 
$I_{t}\simeq I_{b}\;,\;A_{t}\simeq A_{b},\eta \simeq 1$ and Eq.(\ref{19})
implies $\sin 2\varphi _{t}=-\sin 2\varphi _{b}.$ The latter equation has
two possible solutions, namely $\varphi _{t}=-\varphi _{b}$ or $\varphi
_{t}-\varphi _{b}=\pi /2$ $.$ (Of course any shift of the angles by a
multiple of $2\pi $ is also a solution).

\bigskip

a) If $\varphi _{t}=-\varphi _{b}$, $\sigma _{t}=\sigma _{b}$ and from
Eq.(\ref{11}) we obtain $m_{t}=m_{b}$, which is experimentally excluded.

\bigskip

b) If $\varphi _{t}-\varphi _{b}=\pi /2$, Eq.(\ref{16}) leads to $r=1$
and Eqs.(\ref{18}) imply

\begin{equation}
\sin ^{2}\varphi _{t}\simeq \frac{\sigma _{b}^{2}}{\sigma _{t}^{2}+\sigma
_{b}^{2}}\;,\;\sin ^{2}\varphi _{b}\simeq \frac{\sigma _{t}^{2}}{\sigma
_{t}^{2}+\sigma _{b}^{2}}\;.  \label{21}
\end{equation}

\bigskip
\noindent Clearly the large splitting between the physical values of the 
bottom and
top masses ($m_{b}\ll m_{t}$) requires $\sigma _{b}\ll \sigma _{t}$ and thus 
$\varphi _{t}\simeq 0,\varphi _{b}\simeq -\pi /2.$ This in turn demands that
the $CP$-violating phase $\theta $ be close to $\pi /2$ .

To put it differently, the{\em \ presence of a phase} $\theta $ {\em close 
to%
} $\pi /2$ {\em induces both isospin breaking and }$CP${\em \ violation }with

\begin{equation}
\sigma _{b}\ll \sigma _{t}\neq 0\;,\;v\neq 0\;,\;\varphi _{t}\simeq \sigma
_{b}/\sigma _{t}\;,\;\varphi _{b}\simeq -\pi /2+\sigma _{b}/\sigma _{t}\;.
\label{22}
\end{equation}

\medskip
The actual values of the VEVs can be determined from the physical values of
the masses $m_{t},m_{b}$ and $m_{W}.$ As a function of the VEV $v$\ of the
elementary Higgs we find:
\begin{eqnarray}
\sigma _{t}^{2} &=&\frac{\left( 2m_{t}^{2}-h^{2}v^{2}\right) \left(
v_{0}^{2}-v^{2}\right) }{2\left( m_{t}^{2}+m_{b}^{2}-h^{2}v^{2}\right) }\;, 
\nonumber \\
&&  \nonumber \\
\sigma _{b}^{2} &=&\frac{\left( 2m_{b}^{2}-h^{2}v^{2}\right) \left(
v_{0}^{2}-v^{2}\right) }{2\left( m_{t}^{2}+m_{b}^{2}-h^{2}v^{2}\right) }\;, 
\nonumber \\
&&  \nonumber \\
\tan \varphi _{t} &=&\frac{\sigma _{b}}{\sigma _{t}}=\sqrt{\frac{%
2m_{b}^{2}-h^{2}v^{2}}{2m_{t}^{2}-h^{2}v^{2}}}\;,  \label{23}
\end{eqnarray}
\bigskip
and $v$ is determined by the equation
\begin{equation}
\left( m_{t}^{2}+m_{b}^{2}-h^{2}v^{2}-\frac{1}{2}g^{2}\left(
v_{0}^{2}-v^{2}\right) \right) ^{2}=\left( hgv\right) ^{2}\left(
v_{0}^{2}-v^{2}\right) .  \label{24}
\end{equation}
\medskip
Note that Eqs.(\ref{23}) imply an upper bound on $v,$ namely, $v\leq \sqrt{2}%
m_{b}/h.$ This is quite satisfactory since the light quark masses are 
generated by $v$ only.

For small values of $v\ll \sigma _{b},\sigma _{t}$ we have

\begin{eqnarray}
\sigma _{b} &\simeq &\frac{m_{b}v_{0}}{\sqrt{m_{t}^{2}+m_{b}^{2}}}%
\;,\;\sigma _{t}\simeq \frac{m_{t}v_{0}}{\sqrt{m_{t}^{2}+m_{b}^{2}}}\;, 
\nonumber \\
&&  \nonumber \\
\tan \varphi _{t} &\simeq &\frac{m_{b}}{m_{t}}\;,\;g^{2}\simeq \frac{2\left(
m_{t}^{2}+m_{b}^{2}\right) }{v_{0}^{2}}\;.  \label{25}
\end{eqnarray}

For example, with $m_{t}=180$ GeV, $m_{b}=5$ GeV we obtain $\sigma _{b}=6.8$
GeV, $\sigma _{t}=245.9$ GeV, $g=1.03$ and $\varphi _{t}=0.03,$ $\varphi
_{b}=-1.54.$

\medskip

Next we comment on the mass spectrum of the neutral scalars and
pseudoscalars present in our model. To find this spectrum we have to
consider the $6\times 6$ mass matrix given by

\begin{equation}
M_{ij}^{2}=\left. \frac{\partial ^{2}V}{\partial \Phi_{i}\partial \Phi_{j}}%
\right| _{\langle \Phi \rangle },  \label{26}
\end{equation}
\medskip 
where $\Phi _{i}$ denotes any of the real or imaginary parts of the complex
doublets $H^{0},\Sigma _{t}^{0}$ and $\Sigma _{b}^{0}$ .

It is straightforward to find the linear combination of the $\Phi_{i}$'s 
which corresponds to the Goldstone boson eventually eaten up by the $Z^{0}$.
In the $\alpha $-large limit, one of the eigenvalues of the mass matrix will
be proportional to $\sqrt{\alpha \left( \sigma _{b}^{-2}+\sigma
_{t}^{-2}\right) }$ and therefore the corresponding linear combination of
the fields will decouple from the theory. 

The remaining $4\times 4$ mass matrix can be easily diagonalized to obtain 
the other 4 mass eigenstates. We find that the standard Higgs scalar $h$ 
has a mass given by

\begin{equation}
m_{h}\simeq 2g\sqrt{\lambda }m_{t}\;.  \label{27}
\end{equation}
\medskip
For $g\sim 1$ and $\lambda \sim 0.1$, it is of the order of 100 GeV as 
expected \cite{fatelo}. Two of the remaining masses are proportional 
to $\sqrt{\beta }$ and thus quite large. The last mass, $m_{A}$, which 
corresponds mainly to a $\bar{b}\gamma _{5}b$ bound state is very sensitive 
to the difference $h_{t}-h_{b}$. For $h_{t}=h_{b}$ we find 
$m_{A}\simeq 2g\sqrt{\lambda }m_{b}$ at the tree
level, but as soon as $h_{t}$ and $h_{b}$ differ (as expected from higher
order corrections) $m_{A}$ also gets a contribution proportional to 
$\sqrt{\beta }$.

The spectrum of the charged (pseudo) scalar sector can be similarly
analized: apart from the usual $W^{\pm }$ Goldstone bosons, we find two
pairs of complex-conjugate charged bosons, all with a mass proportional 
to $\sqrt{\beta }$.

\section{$CP$ violation}

We have seen that the origin of $CP$ violation in the present model is in
the new interaction characterized by a $\theta \neq 0$ term. This $CP$%
-violating effect filters down to the standard model only if $%
m_{b}/m_{t}\neq 0.$ Let us now investigate whether this
new source of $CP$ violation can be responsible for what is observed in 
the $K^{0}-\bar{K}^{0}$ system.

Let us consider the $3\times 3$ quark mass matrices
\begin{eqnarray}
M_{u} &=&(h)_{u}v+\left( 
\begin{array}{lll}
0 &  &  \\ 
& 0 &  \\ 
&  & 1
\end{array}
\right) g\sigma _{t}e^{i\varphi _{t}},  \nonumber \\
&&  \label{28} \\
M_{d} &=&(h)_{d}v+\left( 
\begin{array}{lll}
0 &  &  \\ 
& 0 &  \\ 
&  & 1
\end{array}
\right) g\sigma _{b}e^{-i\varphi _{b}},  \nonumber
\end{eqnarray}
with ($h)_{u,d}$ arbitrary real matrices.

If we neglect $O(h^{2}v^{2})$ terms, $(MM^{\dagger })_{u,d}$ are
diagonalized by the following unitary matrices:

\begin{equation}
U_{u}\simeq R_{u}\left( 
\begin{array}{lll}
1 &  &  \\ 
& 1 &  \\ 
&  & e^{-i\varphi _{t}}
\end{array}
\right) \;,\;U_{d}\simeq R_{d}\left( 
\begin{array}{lll}
1 &  &  \\ 
& 1 &  \\ 
&  & e^{i\varphi _{b}}
\end{array}
\right) \;,  \label{29}
\end{equation}
with $R_{u,d}$ orthogonal. In this approximation, the
Cabibbo-Kobayashi-Maskawa mixing matrix reads

\begin{equation}
V\equiv U_{u}U_{d}^{\dagger }\simeq R_{u}\left( 
\begin{array}{lll}
1 &  &  \\ 
& 1 &  \\ 
&  & e^{-i(\varphi _{t}+\varphi _{b})}
\end{array}
\right) R_{d}^{T}\;.  \label{30}
\end{equation}

Using the Kobayashi-Maskawa parametrization \cite{kobayashi}

\begin{equation}
V_{KM}\equiv R_{23}(\vartheta _{2})R_{12}(\vartheta _{1})\left( 
\begin{array}{lll}
1 &  &  \\ 
& 1 &  \\ 
&  & e^{i\delta _{KM}}
\end{array}
\right) R_{23}(\vartheta _{3})\;,  \label{31}
\end{equation}
where $R_{ij}(\vartheta )$ denotes a rotation in the $(i,j)$ plane by an
angle $\vartheta ,$ it is obvious that  $\vartheta _{1}$ is arbitrary, $%
\vartheta _{2}=\vartheta _{2}(v/m_{t}),\vartheta _{3}=\vartheta _{3}(v/m_{b})
$ and, last but not least,

\begin{equation}
\delta _{KM}\simeq -(\varphi _{t}+\varphi _{b})\simeq \pi /2\;.  \label{32}
\end{equation}

\medskip The nowadays standard parametrization \cite{PDG} of the CKM mixing
matrix is given by
\begin{equation}
V_{standard}\equiv R_{23}(\vartheta _{23})\left( 
\begin{array}{lll}
1 &  &  \\ 
& 1 &  \\ 
&  & e^{i\delta _{13}}
\end{array}
\right) R_{13}(\vartheta _{13})\left( 
\begin{array}{lll}
1 &  &  \\ 
& 1 &  \\ 
&  & e^{-i\delta _{13}}
\end{array}
\right) R_{12}(\vartheta _{12})\;.  \label{33}
\end{equation}
The $CP$-violating phase $\delta _{13}$ is related to $\delta _{KM}$ by

\begin{equation}
\sin \delta _{13}\simeq \frac{\vartheta _{2}}{\vartheta _{23}}\sin \delta
_{KM}\;,  \label{34}
\end{equation}
in the small mixing angles approximation \cite{fritzsch}. From $%
|V_{ij}|_{KM}=|V_{ij}|_{standard}$ we have
\begin{eqnarray}
\vartheta _{12} &\sim &\vartheta _{1}\;,\;  \nonumber \\
\vartheta _{13} &\sim &\vartheta _{1}\vartheta _{3}\;,\;  \label{35} \\
\vartheta _{23} &\sim &\left( \vartheta _{2}^{2}+\vartheta
_{3}^{2}+2\vartheta _{2}\vartheta _{3}\cos \delta _{KM}\right) ^{1/2}. 
\nonumber
\end{eqnarray}
In particular, we predict
\begin{equation}
\vartheta _{23}\sim \left( \vartheta _{2}^{2}+\vartheta _{3}^{2}\right)
^{1/2}  \label{36}
\end{equation}
if $\delta _{KM}\simeq \pi /2.$ On the other hand, the experimental
constraints \cite{PDG} on the Cabibbo angle $\vartheta _{1}\sim 0.22$ and 
the ratio
\begin{eqnarray}
\left| \frac{V_{ub}}{V_{cb}}\right| &=&0.08\pm 0.02  \nonumber \\
&=&\frac{\vartheta _{13}}{\vartheta _{23}}\simeq 0.22\frac{\vartheta _{3}}{%
\vartheta _{23}}  \label{37}
\end{eqnarray}
require a texture for the $h$-matrices such that $\vartheta _{2}>\vartheta
_{3}.$ From Eqs.(\ref{36}) and (\ref{37}) we obtain
\begin{equation}
\vartheta _{23}\sim \vartheta _{2}\;.  \label{38}
\end{equation}

Therefore, we conclude that our solution leads indeed to a sizeable 
$CP$-violating phase
\begin{equation}
\delta _{13}\simeq \pi /2\;,  \label{39}
\end{equation}
which is welcome by phenomenology in $K^{0}-\bar{K}^{0}$ physics \cite{buras}.

\medskip

To conclude let us summarize the main point of this note: we assume that
there exists a new interaction to which only the third generation of 
quarks ($b$ and $t$) participates. This new interaction is completely 
symmetric in $t$ and $b$ but is supposed to generate a $\theta$ term 
which we take as the
unique source of $CP$ violation. With these assumptions we have shown in a
simple effective model that $\theta $ triggers the breaking of the isospin
symmetry between $t$ and $b$ as expected from general theorems \cite{vafa}.
Self consistency in the context of our model fixes $\theta $ to be around 
$\pi /2$. Due to the smallness of the mass ratio $m_{b}/m_{t}$, this in turn
implies a large $CP$-violating phase in the CKM matrix as required
phenomenologically. Other effects of this new $\theta$ were not considered.

\medskip

\end{document}